# Deep Learning-Driven Segmentation of Ischemic Stroke Lesions Using Multi-Channel MRI

[1]Ashiqur Rahman, [2,*]Muhammad E. H. Chowdhury, [1]Md Sharjis Ibne Wadud, [3]Rusab Sarmun, [4]Adam Mushtak, [4]Sohaib Bassam Zoghoul, [4]Israa Al-Hashimi

[1]Department of Biomedical Physics and Technology, University of Dhaka, Bangladesh. Email: ashiq@bmpt.du.ac.bd (AR); sharjis@du.ac.bd (MSIW)
[2]Department of Electrical Engineering, Qatar University, Doha, Qatar. Email: mchowdhury@qu.edu.qa
[3]Department of Electrical and Electronic Engineering, University of Dhaka, Dhaka 1000, Bangladesh. Email: rusab-2016315002@eee.du.ac.bd (RS)
[4]Department of Radiology, Hamad Medical Corporation, Doha, Qatar. Email: Amushtak@hamad.qa (AM), sohaibzoghoul@gmail.com (SBZ), ialhashimi@hamad.qa (IA)

*Corresponding author: mchowdhury@qu.edu.qa

**Abstract**
Ischemic stroke, caused by cerebral vessel occlusion, presents substantial challenges in medical imaging due to the variability and subtlety of stroke lesions. Magnetic Resonance Imaging (MRI) plays a crucial role in diagnosing and managing ischemic stroke, yet existing segmentation techniques often fail to accurately delineate lesions. This study introduces a novel deep learning-based method for segmenting ischemic stroke lesions using multi-channel MRI modalities, including Diffusion Weighted Imaging (DWI), Apparent Diffusion Coefficient (ADC), and enhanced Diffusion Weighted Imaging (eDWI). The proposed architecture integrates DenseNet121 as the encoder with Self-Organized Operational Neural Networks (SelfONN) in the decoder, enhanced by Channel and Space Compound Attention (CSCA) and Double Squeeze-and-Excitation (DSE) blocks. Additionally, a custom loss function combining Dice Loss and Jaccard Loss with weighted averages is introduced to improve model performance. Trained and evaluated on the ISLES 2022 dataset, the model achieved Dice Similarity Coefficients (DSC) of 83.88% using DWI alone, 85.86% with DWI and ADC, and 87.49% with the integration of DWI, ADC, and eDWI. This approach not only outperforms existing methods but also addresses key limitations in current segmentation practices. These advancements significantly enhance diagnostic precision and treatment planning for ischemic stroke, providing valuable support for clinical decision-making.

**Keywords:** Ischemic stroke segmentation; MRI modalities; deep learning; SelfONN; DWI

## 1. Introduction

Ischemic stroke is a severe and potentially fatal cerebrovascular disorder, marked by the abrupt cessation of blood flow to brain tissue, leading to cellular death and compromised neurological function [1]. The illness poses considerable difficulties in identification and therapy owing to the varied characteristics of ischemia lesions and their intricate morphological alterations [2]. The diagnosis of ischemic stroke remains challenging, with timely and accurate identification of lesions being crucial for effective intervention and improved outcomes [3], [4]. Magnetic Resonance Imaging (MRI) is the gold standard for diagnosing ischemic stroke, offering high-resolution images that provide detailed insights into the extent and severity of brain tissue damage [5], [6]. MRI's ability to visualize various tissue contrasts makes it indispensable for both initial diagnosis and monitoring the progression or response to treatment [7], [8]. However, the variability in lesion appearance ranging from acute infarctions to chronic infarcted areas—presents significant challenges for accurate segmentation and interpretation [9]. This heterogeneity can be ascribed to variations in patient anatomy, the timing of the scan in relation to the stroke event, and the existence of additional diseases [10], [11].

Deep learning technologies have completely changed medical imaging, especially in terms of improving picture segmentation accuracy [12], [13]. Convolutional Neural Networks (CNNs), in particular, are deep learning models that have shown remarkably effective in a variety of medical imaging applications by utilizing enormous datasets and advanced architectures to identify intricate patterns [14]. Segmenting MRI images to delineate ischemic lesions is critical for assessing the extent of stroke, planning treatment, and predicting prognosis [15]. However, the efficacy of these computational approaches hinges on their ability to accurately distinguish and segment the diverse features of ischemic lesions, which can be challenging due to the presence of background noise, motion artifacts, and similarities between lesions and normal brain structures [16].

This study seeks to overcome significant research deficiencies in the segmentation of ischemia lesions from MRI images via deep learning methodologies. We propose a revolutionary deep learning system that addresses the complexity of ischemia lesion segmentation by advancements in models, representations, and methodologies. Prior works in this domain have exhibited notable limitations, such as inadequate handling of background noise and motion artifacts, and the difficulty in distinguishing lesions from normal brain structures. While previous research often focuses on enhancing feature extraction and localization, these approaches may fail to fully mitigate model confusion caused by extraneous elements in the images.
The integration of a DenseNet121 encoder with a Self-Organized Operational Neural Networks (SelfONN) Channel and Space Compound Attention (CSCA) U-Net decoder along with Double Squeeze-and-Excitation (DSE) block, within our segmentation model architecture allows for a nuanced understanding of both local and global image features, enhancing segmentation accuracy. The DenseNet121 encoder is designed to efficiently capture and propagate features through densely connected layers, which helps in retaining critical information across the network.

The Cross-Layer Feature Fusion (CLFF) technique integrates features from various hierarchical levels of the decoder, allowing the model to acquire a more comprehensive and enriched representation of the picture data. Our method integrates information across many scales, thereby capturing nuanced details that conventional single-scale analysis may miss. This hierarchical feature fusion enables the model to utilize both high-level semantic information and low-level spatial data, resulting in enhanced segmentation accuracy.

We also leverage multimodal learning by combining Diffusion-Weighted Imaging (DWI) and Apparent Diffusion Coefficient (ADC) modalities into a multi-channel image, enhancing the differentiation and segmentation of ischemic lesions compared to using single modality images. DWI and ADC provide complementary information about the diffusion properties of water molecules in brain tissue, which is crucial for identifying areas affected by ischemia. The varied information from these multiple sources enriches the feature representation, enabling the segmentation algorithm to better understand the visual properties of different lesion regions. This multimodal approach improves the model's ability to detect and segment lesions that might be challenging to identify using a single modality.

The key contributions of our study can be outlined as follows:
- We propose a novel model combining a DenseNet121 encoder with a SelfONN CSCA U-Net decoder, advancing feature learning capabilities through effective feature fusion techniques.
- We leverage multimodal learning by integrating DWI and ADC modalities into a multi-channel image, significantly improving segmentation accuracy across all lesion regions.
- We propose a custom loss function combining Jaccard loss and Dice loss, enhancing the model's performance in handling class imbalance and improving segmentation accuracy.

This article comprises five principal components. Section 2 examines the pertinent literature. Section 3 delineates the research technique employed in this study. Section 4 delineates the results and offers a comprehensive examination of the model's efficacy. Finally, Section 5 closes the work.

## 2. Related Works

This section reviews significant studies on ischemic lesion segmentation using deep learning techniques with a focus on MRI-based approaches. The key aspects considered in the literature include the methodologies, datasets, imaging modalities, results, and limitations of various studies. Gheibi et al. (2023) developed a CNN-Res architecture with a U-shaped structure, incorporating residual units and a bottleneck strategy [17]. Evaluating their approach on datasets from Tabriz University and the SPES 2015 competition using FLAIR and DWI modalities, they achieved DSC of 85.43% and 79.23%, respectively. The study highlighted issues such as small lesion size bias in segmentation and potential gradient vanishing and deepening network problems due to ResNet blocks. Bal et al. (2024) employed various MRI modalities (T1, T2, DWI, and FLAIR) from the ISLES2015 dataset in their pre-processing, patch creation, data augmentation, and CNN classification approach [18]. They reported a DSC of 0.85. However, they noted the limited use of multiple MRI modalities in existing models and the need for more advanced deep learning methods for ischemic stroke lesion segmentation.

Wu et al. (2023) proposed a W-Net model integrating CNN and transformer-based methods for lesion segmentation [19]. Their approach, tested on the ATLAS (with T1, ADC, DWI, and FLAIR) and ISLES2022 datasets, outperformed existing methods with DSC of 61.76 and 76.47, Hausdorff Distance (HD) of 32.47, and F2 scores of 64.60. Challenges included variability in lesion areas due to individual differences and difficulties in generating trusted boundaries for stroke lesions. Alshehri et al. (2023) integrated a few-shot learning strategy with a base CNN model for segmentation using the ISLES 2015 dataset and FLAIR and DWI modalities [20]. With a DSC of 0.68, they were able to identify the difficulties in comparing stroke segmentation methods with the help of the ISLES SISS challenge dataset.

Thiyagarajan et al. (2023) utilized an Arithmetic Optimization-based K-Means (AOK-Means) approach for cluster center optimization [21]. Applying their method to the ISLES2015 dataset with DWI modality, they reported DSC of 70.8% and 76.4%, with precision values of 90.4% and 91.9%. Challenges encompassed the incapacity to identify clusters prior to segmentation and the obstruction posed by random cluster center initialization in attaining a globally optimal solution. Moon et al. (2022) assessed 2D and 3D U-Net convolutional neural network architectures for the segmentation of stroke lesions utilizing brain scans from 79 acute ischemic stroke patients, employing FLAIR and DWI modalities [22]. The optimal outcomes were attained using a 2D multimodal U-Net, resulting in a mean DSC of 0.737. The study encountered constraints in delineating brainstem stroke lesions owing to insufficient data and the difficulty posed by an uneven distribution of lesion sizes.

Jazzar et al (2022) developed a modified U-Net architecture with multi-path integration, achieving a DSC of 0.84 on the ISLES2015 dataset using FLAIR and DWI modalities [23]. They addressed class imbalance through optimized loss functions, improving robustness in segmentation. Platscher et al. (2022) explored the application of generative models, specifically Pix2Pix, CycleGAN, and SPADE, for the purpose of stroke lesion labeling and brain segmentation [24]. Utilizing a dataset comprising 804 cases from the University Hospital Basel and employing the DWI modality, they achieved a DSC of 72.8%. Despite this, their study encountered notable challenges. The use of synthetic data was constrained by data privacy concerns, and the models trained exclusively on this synthetic data demonstrated insufficient competitive performance.

With the use of the DWI modality from the Houston Methodist Hospital Stroke Registry, Wong et al. (2022) presented a deep learning model with rotation-reflection equivariance for stroke volume segmentation [25]. Their model received an 85% Dice score; nevertheless, segmentation performance was impacted by differences in MRI images caused by scan parameters. Using the ISLES 2015 dataset with T1, T2, DWI, and FLAIR modalities, Karthik et al. (2020) suggested a Fully Convolutional Network (FCN) with an attention mechanism for ischemic lesion segmentation [26]. They reported a DSC of 0.75, emphasizing issues such CNNs' inability to provide local context for accurate segmentation and their inability to distinguish between features because of modality differences.

The aforementioned research shows how deep learning methods can enhance the precision and effectiveness of ischemic lesion segmentation from MRI images. Despite notable advancements, challenges such as small lesion size bias, variability in lesion areas, and the need for large, annotated datasets remain. Future research should focus on overcoming these limitations, enhancing model robustness, and exploring the integration of multiple MRI modalities to further improve segmentation performance.

## 3. Experimental Methodology

This section outlines the experimental framework employed to build and assess our proposed model for ischemic lesion segmentation in MRI images. The ISLES 2022 dataset, comprising DWI and ADC modalities, underwent pre-processing to improve lesion visibility and maintain uniform input dimensions. Our proposed model leverages DenseNet121 as the encoder and SelfONN in the decoder, with the integration of CSCA attention mechanisms and DSE blocks to refine feature extraction. CLFF was employed to retain critical details during up-sampling. The model was optimized using a composite loss function combining Dice and Jaccard losses for improved segmentation accuracy. The overall process is depicted in Figure 1, illustrating the various stages from dataset preparation to model evaluation.

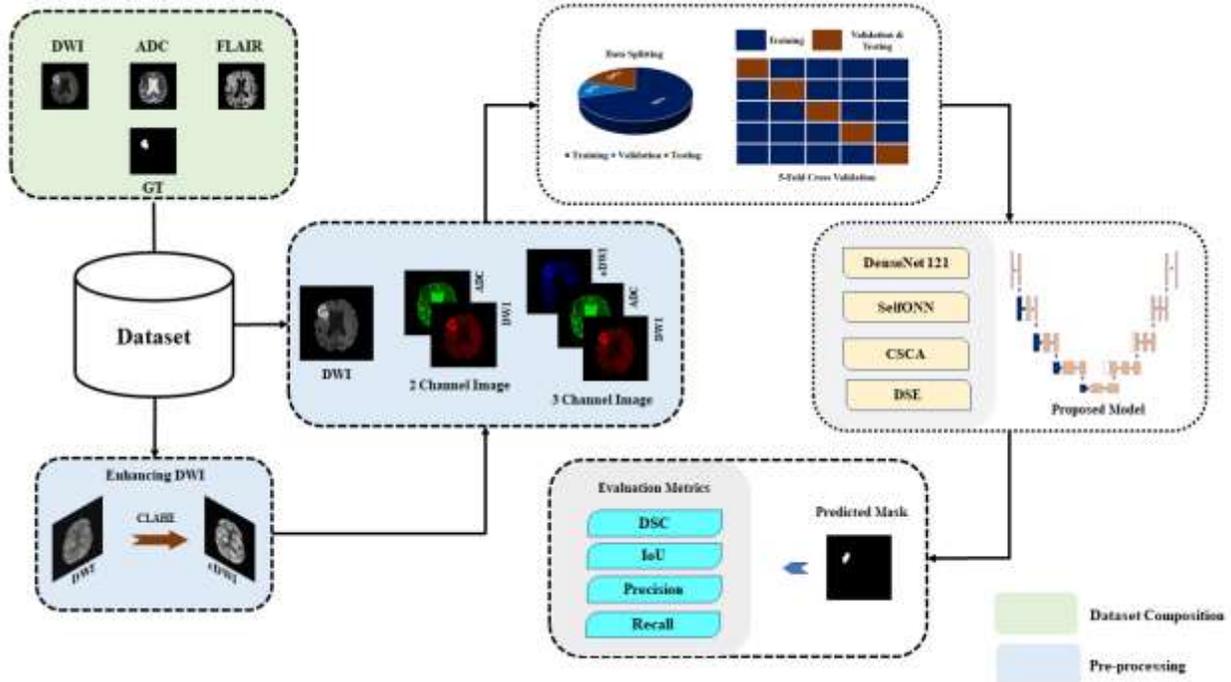

**Figure 1.** Schematic representation of the experimental pipeline for ischemic lesion segmentation.

*3.1 Dataset Descriptions*

The ISLES 2022 dataset is a multi-center collection of MRI scans, meticulously annotated for the segmentation of acute and subacute ischemic stroke lesions [27]. This dataset, central to the ISLES challenge, includes 400 MRI cases from various medical centers, utilizing different MRI machines and protocols. The cases vary in lesion size, location, and ischemic stage. Annotations were performed using a hybrid human-algorithm approach, enhancing precision through expert radiologist review and refinement. The dataset is divided into 250 training cases, publicly available for model development, and 150 testing cases reserved for independent validation. Each training case includes images from three MRI modalities—FLAIR, DWI, and ADC—comprising 52,739 FLAIR slices and 15,684 slices each for DWI and ADC, all paired with segmentation masks. This dataset supports the development of robust models capable of generalizing across diverse patient populations and imaging conditions, ultimately aiding in clinical decision-making and improving patient outcomes.

*3.2 Data Preprocessing*

In this study, we focused on the DWI and ADC modalities of the ISLES 2022 dataset, excluding FLAIR images to concentrate on the most relevant inputs for ischemic lesion segmentation. The MRI scans, initially in Neuroimaging Informatics Technology Initiative (NIfTI) format, were converted to PNG and resized to 256×256 pixels to ensure consistent input dimensions for our deep learning model. To enhance lesion visibility, we applied Contrast Limited Adaptive Histogram Equalization (CLAHE) to the DWI images, improving the contrast and making subtle features more distinguishable. We then created two types of composite images: a 2-channel image combining DWI and ADC, and a 3-channel image incorporating DWI, ADC, and the enhanced DWI. This approach was designed to provide the model with rich and complementary information from the different MRI modalities, ultimately enhancing its ability to accurately segment ischemic lesions by capturing a broader range of critical features.

*3.3 Experimental Details*

To mitigate bias and prevent data leakage, a patient-wise data splitting approach was employed, as opposed to the conventional random image splitting method. This approach ensures that all data pertaining to a single patient is consistently allocated to either the training, validation, or test set. This approach offers a more dependable assessment of the model's ability to generalize to unfamiliar patient data [28]. The dataset, consisting of 15,684 2D slices from 250 patients, was partitioned into training, validation, and testing sets in the ratios of 70%, 10%, and 20%, respectively. Furthermore, 5-fold cross-validation was utilized to provide a rigorous assessment. The computational setup included an Intel Core i9-13900K CPU featuring 24 cores at 3.00 GHz, an NVIDIA GeForce RTX 4090 GPU with 24 GB of VRAM, and 64 GB of DDR5 RAM. The configuration employed CUDA 11.8 and PyTorch 2.3.1 to enable efficient execution of deep learning operations.

*3.4 Our Proposed Model*

Our proposed model, illustrated in Figure 2, introduces a novel approach to ischemic lesion segmentation by leveraging the strengths of DenseNet121 as the encoder and SelfONN in the decoder. Additionally, the model integrates CSCA blocks and DSE blocks to enhance feature extraction and improve segmentation accuracy. This unique combination aims to provide superior performance in medical image segmentation tasks.

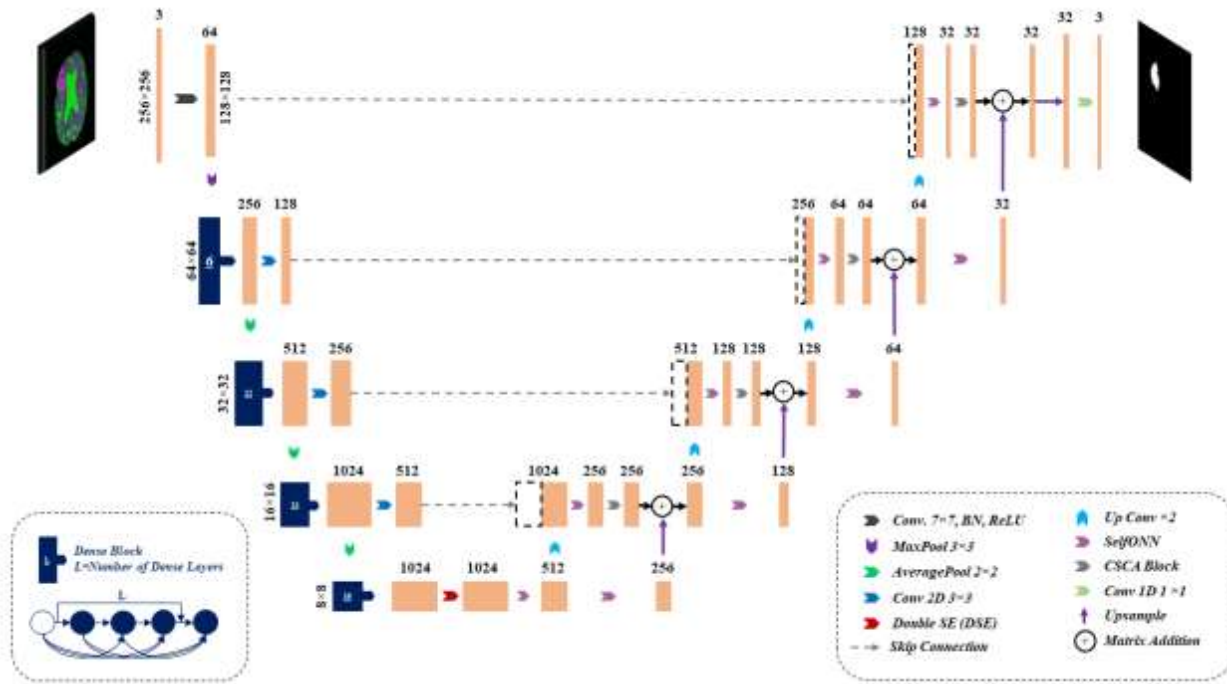

**Figure 2.** Detailed architecture of our proposed model.

*3.4.1 Encoder: DenseNet121*

To achieve efficient and robust feature extraction, we incorporated DenseNet121 as the encoder in our model. Figure 3 illustrates how DenseNet121, known for its dense connectivity through densely connected convolutional blocks, facilitates better gradient flow and feature reuse, which are crucial for medical image segmentation [29]. We utilized the pre-trained DenseNet121 up to its last dense block, excluding the classification layer, to extract rich and detailed feature maps from the input images.

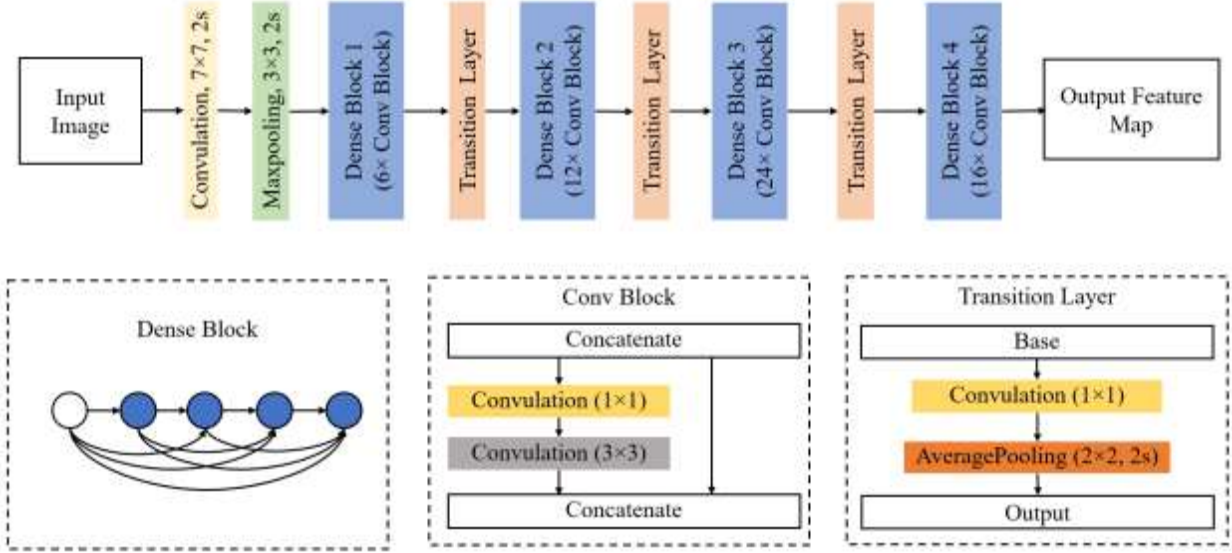

**Figure 3.** The architecture of DenseNet121 was used as the encoder in our model, showcasing its dense connectivity and the extraction of feature maps up to the last dense block.

DenseNet121 organizes its layers into blocks, with each layer directly linked to every other layer in a feed-forward manner. This intricate connectivity structure alleviates the vanishing gradient issue, promotes feature reutilization, and markedly decreases the parameter count relative to conventional convolutional networks. The features obtained via DenseNet121 function as the input for our innovative decoder architecture.

*3.4.2 SelfONN*

Self-ONNs enhance Operational Neural Networks (ONNs) by incorporating generative neurons that adapt and optimize their nodal operators during training [30]. This adaptation results in greater flexibility and computational efficiency, allowing each neuron to generate and optimize a combination of nodal operators.

In conventional CNNs, the input map for the $k - th$ neuron in the current layer, denoted as $x_k^l$, is derived from the following equation:

$$x_k^l = b_k^l + \sum_{i=1}^{N_{l-1}} \text{conv } 2D(w_{ki}^l, y_i^{l-1}) \qquad (1)$$

$$x_k^l(m,n)\Big|_{(0,0)}^{(M-1,N-1)} = \sum_{r=0}^{2} \sum_{t=0}^{2} \left(w_{ki}^l(r,t) y_i^{l-1}(m+r, n+t)\right) + \cdots \qquad (2)$$

ONNs extend this by incorporating nodal and pool operators, described by:

$$x_k^l = b_k^l + \sum_{i=1}^{N_{l-1}} \text{oper } 2D(w_{ki}^l, y_i^{l-1}) \qquad (3)$$

$$x_k^l(m,n)\Big|_{(0,0)}^{(M-1,N-1)} = b_k^l + \sum_{i=1}^{N_{l-1}} \left( P_k^l \begin{bmatrix} \Psi_{ki}^l(w_{ki}^l(0,0), y_i^{l-1}(m,n)), \dots, \\ \Psi_{ki}^l(w_{ki}^l(r,t), y_i^{l-1}(m+r, n+t)), \dots \end{bmatrix} \right) \quad (4)$$

Self-ONNs further improve upon ONNs by using a composite nodal operator that is iteratively created during backpropagation. This composite operator is expressed as a Q-th order Taylor approximation:

$$\Psi(w, y) = w_0 + w_1 y + w_2 y^2 + \dots + w_Q y^Q \quad (5)$$

During forward propagation, each kernel element's nodal operator in Self-ONNs is approximated by this composite nodal operator:

$$\Psi(w, y_{(m+r,n+t)}) = w_{r,t,0} + w_{r,t,1} y_{(m+r,n+t)} + w_{r,t,2} y^2_{(m+r,n+t)} + \dots + w_{r,t,Q} y^Q_{(m+r,n+t)} \quad (6)$$

This allows neurons to self-organize their nodal operators during training, optimizing their functions for maximum learning performance. Experimental results show that Self-ONNs outperform conventional ONNs and CNNs in both learning capability and computational efficiency, even with more compact networks.

### 3.4.3 Channel and Space Compound Attention (CSCA)

We integrate CSCA mechanism is designed to enhance feature representations by integrating both channel and spatial attention mechanisms [31]. The detailed architecture of the CSCA mechanism is illustrated in Figure 4. The CSCA mechanism starts with an input feature map that is processed along three distinct paths.

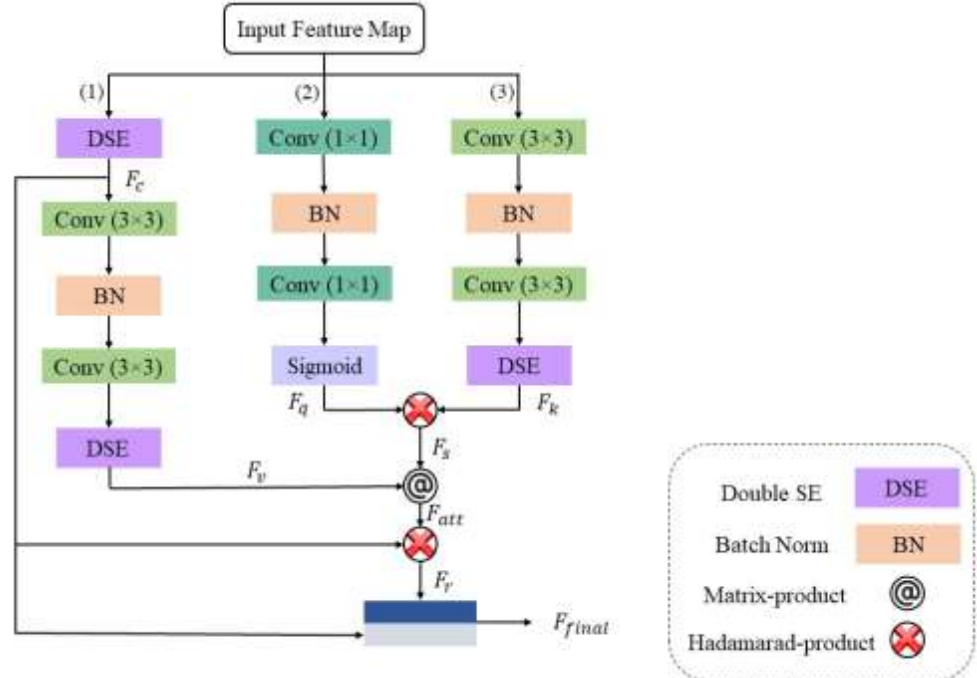

**Figure 4.** The structure of the CSCA mechanism. The input feature map is processed through three paths to generate channel-wise ($F_v$) and spatial-wise ($F_q$ and $F_k$) attention features. These are combined to produce the final attention-enhanced feature map ($F_{final}$).

In path (1), the input feature map is passed through a DSE module, which emphasizes important channel-wise features and produces an intermediate feature map $F_c$. Subsequently, this map undergoes further refinement via a series of convolutional layers. A 3x3 convolution is applied, followed by batch normalization (BN), and a further 3x3 convolution layer generates the final channel-wise attention feature map $F_v$.

In path (2), the input feature map is analyzed to provide spatial attention features. The input feature map undergoes initial processing using a 1x1 convolution layer, succeeded by batch normalization. This is succeeded by an additional 1x1 convolutional layer. A sigmoid activation function is subsequently used, generating a spatial attention map $F_q$.

In path (3), the input feature map undergoes another spatial processing sequence. The input feature map is processed through a 3x3 convolution layer followed by BN, and then another 3x3 convolution layer. This feature map is further refined through a DSE module, emphasizing crucial spatial features and producing the final spatial-wise feature map $F_k$.

Finally, the outputs from these paths are combined. The channel-wise feature map $F_v$ is combined with the spatial attention map $F_s$ (obtained by multiplying $F_q$ with $F_k$ elementwise) through a matrix product operation, resulting in the intermediate feature map $F_{att}$. The input feature map also generates $F_r$ through another path, which is then multiplied elementwise with $F_{att}$ to produce the final attention-enhanced feature map $F_{final}$. The CSCA mechanism improves the model's focus on key features by incorporating both channel and spatial attention, leading to enhanced performance in tasks such as segmentation and classification.

*3.4.4 Double Squeeze-and-Excitation (DSE)*

We incorporated a channel attention mechanism called the DSE block in the bottleneck layer to further enhance high-level semantic features. The DSE block adjusts channel weights to better capture important details in the input features by utilizing global average pooling (GAP) and global maximum pooling (GMP) [32]. The structure of the DSE block is illustrated in Figure 5.

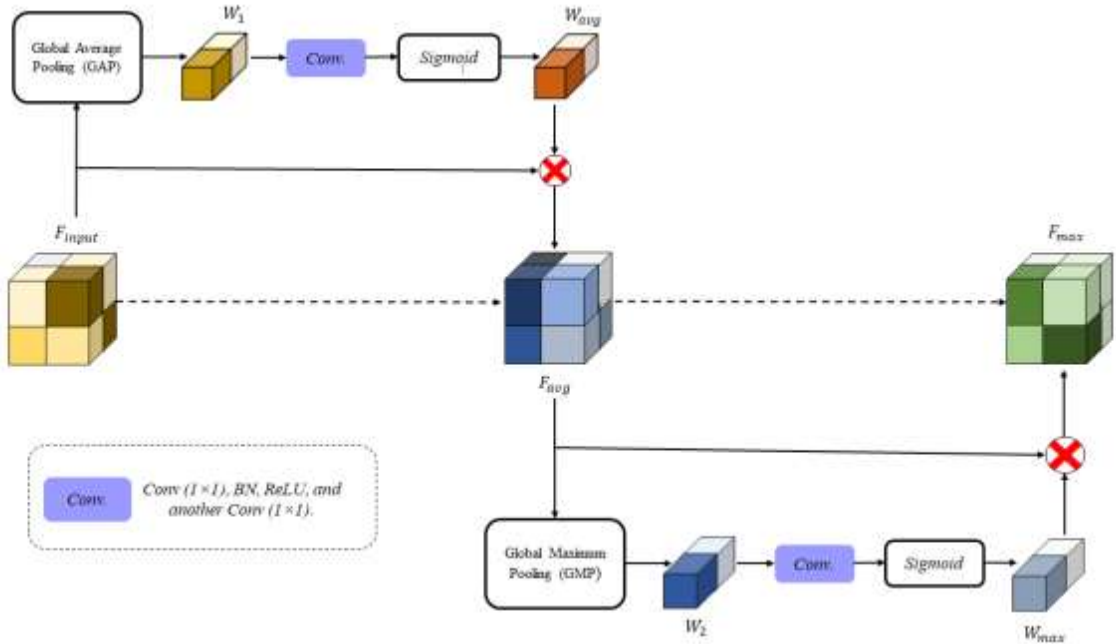

**Figure 5.** Architecture of the Double Squeeze-and-Excitation (DSE) Block: The DSE block incorporates both Global Average Pooling and Global Maximum Pooling to create attention weights, refined through convolutional blocks. These weights enhance the feature maps through element-wise multiplication, improving the network's focus on key features.

First, the input feature map $F_{input}$ undergoes global average pooling to produce a weight vector $W_{avg}$.

The weight vector is subsequently fed into a convolutional block, which begins with a 1 × 1 convolution, followed by batch normalization ($BN$) and $ReLU$ activation, and is completed by another 1 × 1 convolution, with a sigmoid activation applied at the end to produce the normalized weight vector $W_1$. The feature map $F_{avg}$ is generated by hadamard product between $F_{input}$ and $W_1$.

Next, the feature map $F_{avg}$ is transformed using global maximum pooling to produce another weight vector $W_{max}$. This weight vector is also processed through the *Conv.* block, and a sigmoid activation function to yield the normalized weight vector $W_2$. The final output feature map $F_{max}$ is obtained by performing an element-wise multiplication between $F_{avg}$ and $W_2$.

By applying both GAP and GMP, the DSE block creates attention maps that effectively emphasize the most informative features in the input feature maps. This attention mechanism enhances the overall performance of the neural network by focusing on both average and maximum pooled features, leading to a more robust and discriminative feature representation.

### 3.4.5 Cross-Layer Feature Fusion (CLFF)

To address the loss of information during up-sampling, we employed CLFF in the decoder. The CLFF combines features from different layers, ensuring the retention and effective merging of crucial details [33]. This technique aids in recovering lost information, thereby contributing to the improved segmentation accuracy of the model. Mathematically, CLFF can be represented as:

$$F_{fused} = \sigma(W_i * F_i + W_j * F_j) \qquad (7)$$

where $F_i$ and $F_j$ are the feature maps from layers $i$ and $j$, respectively, $W_i$ and $W_j$ are learnable weight matrices applied to $F_i$ and $F_J$, σ is an activation function, and ∗ denotes the convolution operation. This fused feature map $F_{fused}$ is then utilized in subsequent layers of the decoder, enhancing segmentation performance by integrating multi-scale contextual information.

### 3.4.6 Loss Function

In this study, we proposed a composite loss function by integrating Dice and Jaccard losses to improve the segmentation model's performance. Dice Loss, based on the Dice coefficient, quantifies the overlap between the predicted segmentation and the actual labels, mitigating class imbalance by emphasizing the regions of interest [34]. The Dice coefficient D is defined as:

$$D = \frac{2\sum_i(p_i g_i)}{\sum_i(p_i) + \sum_i(p_i) + \epsilon} \qquad (8)$$

where $p_i$ is the predicted probability for pixel $i$, $g_i$ is the ground truth label for pixel $i$, and $\epsilon$ is a small constant added for numerical stability. Dice Loss $L_d$ is then calculated as:

$$L_d = 1 - D \qquad (9)$$

Jaccard Loss, based on the Jaccard Index, similarly quantifies the similarity between the predicted and actual segmentations, ensuring accurate boundary predictions [35]. The Jaccard Index $J$ is defined as:

$$J = \frac{2\sum_i(p_i g_i)}{\sum_i(p_i + g_i - p_i g_i) + \epsilon} \qquad (10)$$

and the Jaccard Loss $L_j$ is calculated as:

$$L_j = 1 - J \qquad (11)$$

By combining these two loss functions, we leverage their respective strengths to achieve superior segmentation performance. The composite loss function is defined as a weighted sum of Dice Loss and Jaccard Loss:

$$L = 0.5 \cdot L_d + 0.5 \cdot L_j \tag{12}$$

*3.4.7 Evaluation Metrics*

To assess the efficacy of our segmentation model, we employed several critical evaluation metrics: Dice Similarity Coefficient (DSC), Intersection over Union (IoU), Precision, and Recall.

The Dice Similarity Coefficient (DSC) quantifies the overlap between the ground truth and the anticipated segmentation. It is delineated as:

$$DSC = \frac{2 \times TP}{2 \times TP + FP + FN} \tag{13}$$

The IoU quantifies the similarity between the predicted and ground truth segmentations. It is calculated as:

$$IoU = \frac{TP}{TP + FP + FN} \tag{14}$$

Precision (P) measures the accuracy of positive predictions. It is given by:

$$Precision = \frac{TP}{TP + FP} \tag{15}$$

Recall (R), also known as Sensitivity, assesses the model's ability to identify all relevant instances. It is defined as:

$$Recall = \frac{TP}{FN + TP} \tag{16}$$

where $TP$ are True Positives, $FP$ are False Positives, and $FN$ are False Negatives.

## 4. Results and Discussion

This section presents the results of our experiments and provides an in-depth discussion of the findings. We evaluate the performance of the proposed model against baseline models, assess its effectiveness on different imaging modalities, and analyze the impact of various loss functions on segmentation accuracy. The results are quantified using standard performance metrics and illustrated through qualitative examples.

*4.1 Comparison with Baseline Models*

Our proposed model outperforms several of the best-performing baseline models, including DenseNet121_UNet, DenseNet121_UNet++, and DeepLabV3, as demonstrated in Table 1. For training and evaluation, we utilized 3-channel input images consisting of DWI, ADC, and enhanced DWI modalities. The model achieved the highest Dice Similarity Coefficient (87.49%) and Intersection over Union (84.63%), indicating superior accuracy and overlap with the ground truth. While the precision (94.87%) is slightly lower than that of DenseNet121_UNet (95.01%), our model's higher recall (88.57%) reflects a better balance between detecting true positives and minimizing false negatives.

**Table 1.** Our proposed model's performance compared to baseline models.

| Model | DSC | IoU | Precision | Recall |
|---|---|---|---|---|
| **DenseNet121_UNet** | 86.43 | 83.50 | 95.01 | 87.15 |
| **DenseNet121_UNet++** | 86.35 | 83.41 | 94.43 | 87.61 |
| **DeepLabV3** | 84.40 | 81.38 | 93.60 | 85.40 |
| **Ours Proposed Model** | 87.49 | 84.63 | 94.87 | 88.57 |

Figure 6 visually supports these findings, showing that the segmentation results from our proposed model align closely with the ground truth. Compared to the baseline models, our model demonstrates fewer false positives and false negatives. DenseNet121a-UNet and UNet++ tend to produce more false positives, while DeepLabV3 struggles with both false positives and false negatives, especially in smaller lesions. Overall, our model demonstrates a robust and effective approach to ischemic lesion segmentation, significantly improving upon existing methods.

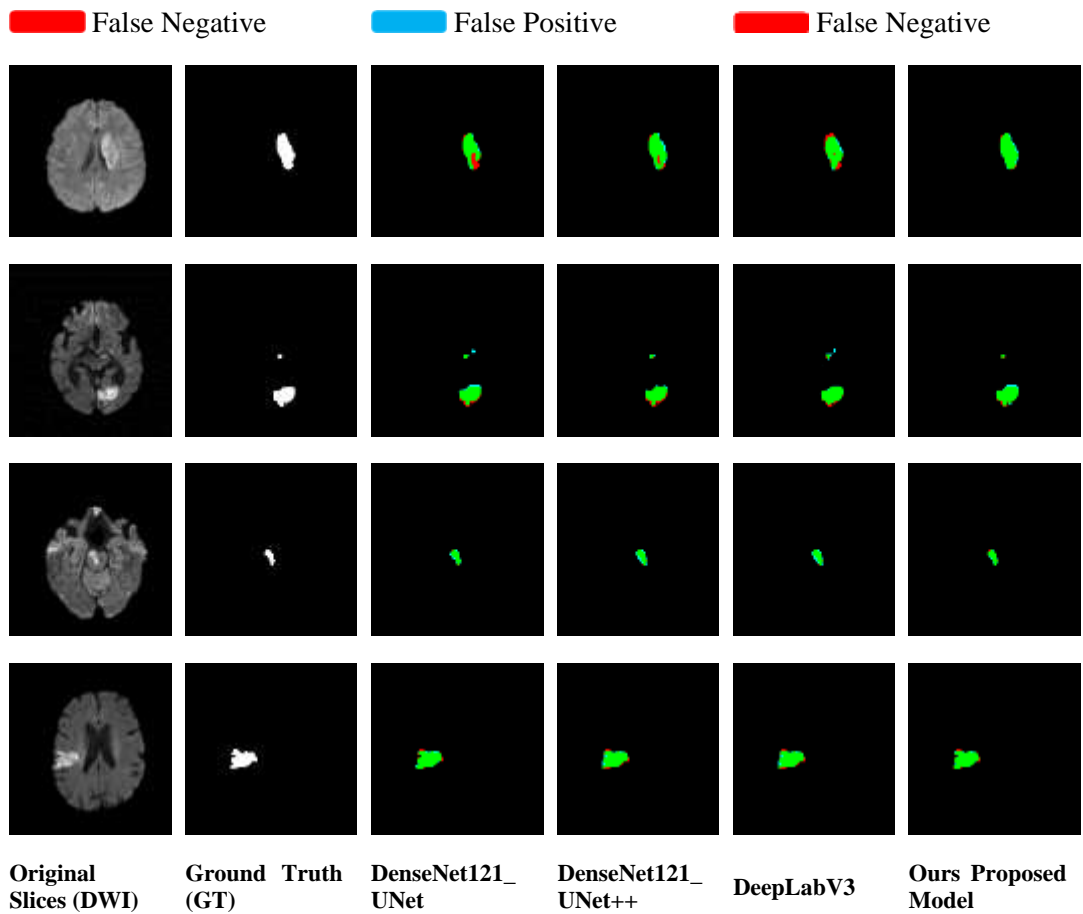

**Figure 6.** Our proposed model's qualitative performance compared to baseline models.

*4.2 Comparison Multimodal Learning*

We assessed the performance of our proposed ischemic lesion segmentation model with three distinct input setups: (1) using only DWI, (2) combining DWI and ADC, and (3) integrating DWI, ADC, and enhanced DWI.

**Table 2.** Performance comparison of our proposed model with different input modalities

| Modality | DSC | IoU | Precision | Recall |
|---|---|---|---|---|
| **DWI** | 83.88 | 81.09 | 95.48 | 84.29 |
| **DWI + ADC** | 85.86 | 82.95 | 96.40 | 85.47 |
| **DWI + ADC + eDWI** | 87.49 | 84.63 | 94.87 | 88.57 |

The quantitative results, summarized in Table 2, show that using DWI alone provided a strong baseline, achieving a DSC of 83.88% and an IoU of 81.09%. Incorporating ADC alongside DWI slightly improved the segmentation accuracy, increasing the DSC to 85.86% and IoU to 82.95%. The most significant enhancement was observed when eDWI was added, resulting in a DSC of 87.49% and an IoU of 84.63%, indicating better alignment with the ground truth.

The qualitative results, presented in Figure 7, further illustrate these findings. The segmentation maps show that the DWI-only configuration often under-segments lesions, particularly smaller or more subtle regions. When ADC is added (2 channel), the segmentation becomes more thorough, capturing areas that were previously missed. However, this configuration introduces some minor artifacts, as evidenced by the occasional over-segmentation in non-lesion areas. The inclusion of enhanced DWI leads to a clearer and more accurate delineation of lesion boundaries, reducing false positives and resulting in more precise and consistent segmentation outcomes.

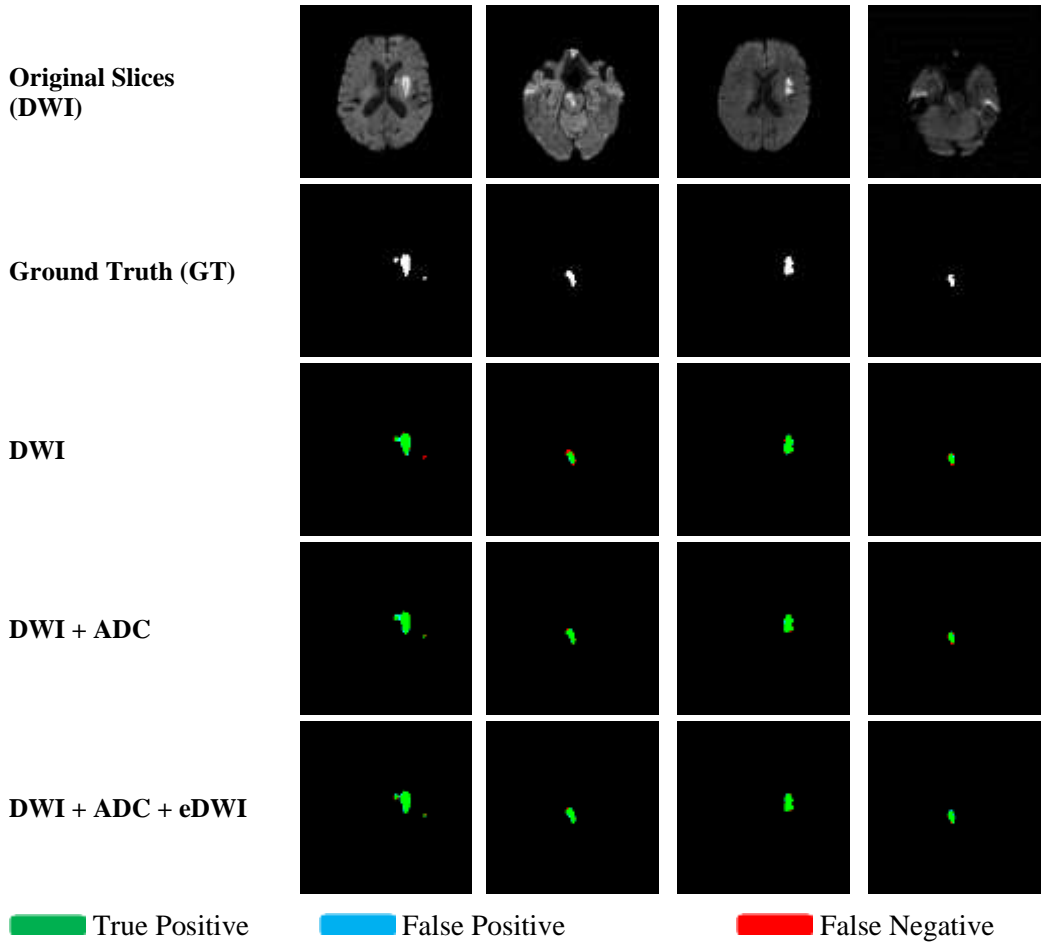

**Figure 7.** Qualitative performance comparison of our proposed model with different input modalities.

*4.3 Comparison of Loss Functions*

The proposed combined loss function, which integrates Dice Loss and Jaccard Loss, demonstrated superior performance in ischemic lesion segmentation compared to using Dice or Jaccard Loss alone. As shown in

**Table 3**, The Combined Loss function achieved the highest DSC of 87.49% and the highest IoU of 84.63%, indicating better segmentation accuracy and overlap with the ground truth. It also delivered the highest precision at 94.87%, with a slight trade-off in recall (88.57%) compared to Dice Loss (88.75%). This balance between precision and recall suggests that the combined loss effectively reduces both false positives and false negatives.

**Table 3.** Comparison of our proposed model's performance using various loss functions.

| Loss Function | DSC | IoU | Precision | Recall |
|---|---|---|---|---|
| **Dice Loss** | 86.52 | 83.47 | 93.29 | 88.75 |
| **Jaccard Loss** | 86.47 | 83.53 | 94.47 | 87.67 |
| **Combined Loss** | 87.49 | 84.63 | 94.87 | 88.57 |

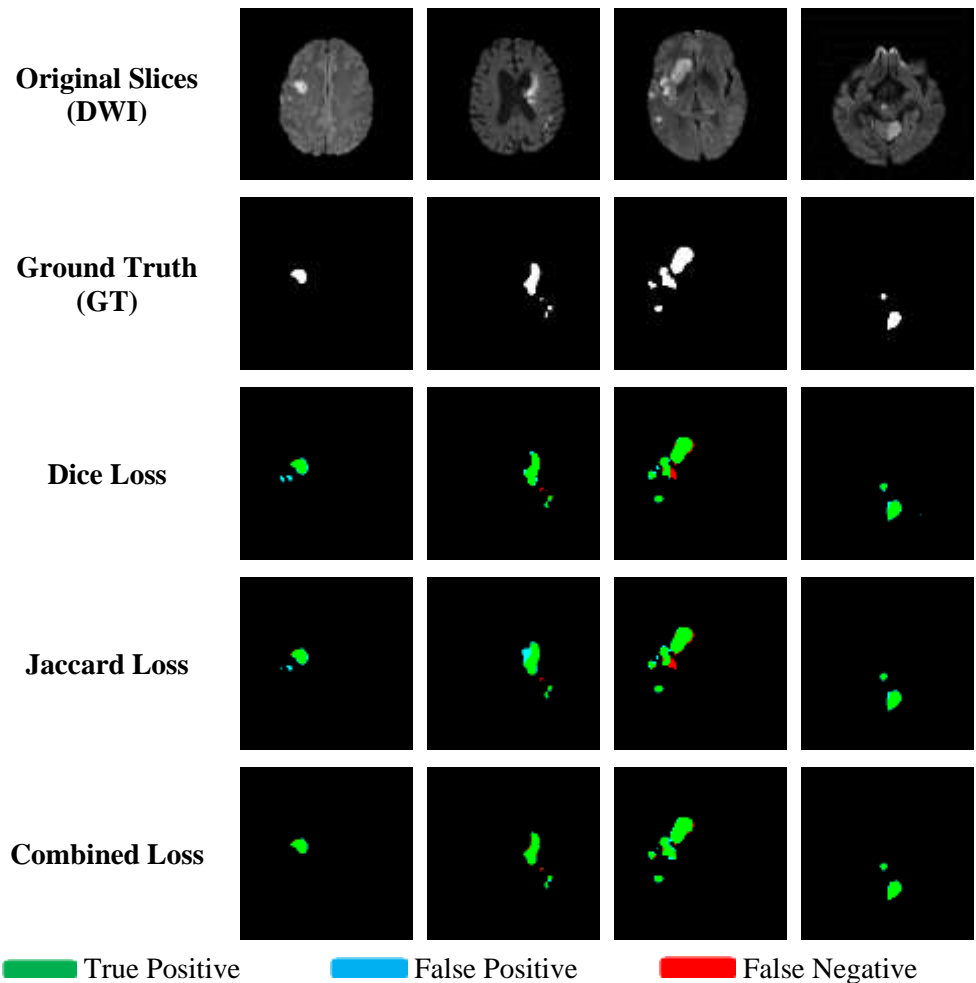

**Figure 8**. Qualitative performance comparison of our proposed model with different loss functions.

Figure 8 illustrates the segmentation results, where the outputs using the Composite Loss closely match the ground truth, with fewer instances of over-segmentation and under-segmentation. In contrast, Dice Loss shows a tendency to over-segment, leading to more false positives, while Jaccard Loss, although slightly better in this regard, struggles with capturing smaller lesion regions. The Combined Loss successfully merges the sensitivity of Dice Loss with the rigor of Jaccard Loss, resulting in a more accurate and reliable segmentation outcome. This innovative approach offers a robust solution for enhancing ischemic lesion segmentation performance.

*4.4 Comparison with State-of-the-Art Models*

We compared our segmentation results against the top-performing models from the ISLES 2022 Challenge, which serves as a benchmark for ischemic stroke lesion segmentation using modalities like DWI, ADC, and FLAIR modalities [36]. The top three models in the challenge were developed by teams SEALS, NVAUTO, and SWAN. The SEALS team's model, utilizing a 3D nnUNet architecture with DWI and ADC inputs, a dice loss, and categorical cross-entropy, produced a mean DSC of 82.0%. The model by team NVAUTO reached a mean DSC of 82.0% as well, employing a SegResNet-based solution with various data augmentation techniques and a Dice-Focal loss function [37]. The created by SWAN team obtained a mean DSC of 81.0%, utilizing a Swin Transformer architecture integrated with Non-negative Matrix Factorization (NMF) layers [38]. In contrast, our Model achieved a significantly higher DSC of 87.4%, representing an improvement of 5.4% over the SEALS and NVAUTO models and 6.4% over the SWAN model, as summarized in Table 4.

**Table 5**. Performance of our proposed model compared to State-of-The-Art models.

| Model | Architecture | Input Modalities | Loss Function | DSC |
|---|---|---|---|---|
| SEALS | nnUNet | DWI, ADC | Dice + Cross-Entropy | 0.82 |
| NVAUTO | SegResNet | DWI, ADC | Dice + Focal Loss | 0.82 |
| SWAN | Swin Factorizer | DWI, ADC, FLAIR | Dice + Cross-Entropy | 0.81 |
| Proposed Model | SelfONN-CSCA-UNet | DWI, ADC | Jaccard + Dice | 0.8749 |

## 5. Limitations and Directions for Future Research

Although this study has shown strong performance in segmenting ischemic stroke lesions, it has several notable limitations. The study relied on the ISLES 2022 dataset, which is focused specifically on ischemic stroke lesions. However, in real-world clinical scenarios, patients may present with various neurological conditions beyond ischemic strokes, such as tumors, hemorrhages, or other brain pathologies that might exhibit similar imaging characteristics. For instance, certain brain tumors or hemorrhagic lesions could mimic the appearance of ischemic stroke lesions due to overlapping signal characteristics on MRI scans [39]. This limitation highlights that our model's ability to accurately differentiate ischemic strokes from other brain conditions could not be thoroughly evaluated, raising concerns about its generalizability in clinical settings where multiple conditions are present simultaneously. Another drawback of this study is the exclusive use of DWI, ADC, and enhanced DWI modalities. Although these modalities are crucial for identifying ischemic strokes, the exclusion of other imaging techniques, such as FLAIR or T2-weighted images, could limit the model's robustness [40]. In clinical settings, these additional modalities offer crucial insights into lesion characteristics, and their exclusion from the current study may limit the model's effectiveness in more complex diagnostic situations. Additionally, the model sometimes struggles with accurately segmenting very small infarcts. Small infarcts can be challenging to detect and delineate due to their subtle appearance on MRI scans and the limited spatial resolution of the imaging modalities used [41]. This limitation is particularly concerning in clinical practice, where the accurate identification of even the smallest infarcts can be critical for timely and appropriate treatment.

To better align with the complexities of real-world clinical needs, future research will focus on expanding the dataset to include a broader range of brain pathologies, such as tumors, hemorrhages, and other conditions. This expansion will enable the model to learn and accurately differentiate ischemic stroke lesions from other brain conditions, thereby improving its generalizability and clinical applicability. In addition, future work should explore the inclusion of other imaging modalities like FLAIR and T2-weighted images. These modalities could enhance the model's robustness and its ability to detect and analyze lesions with greater accuracy. Expanding the scope of the model to integrate these additional modalities will move us toward developing a more comprehensive and versatile tool that can serve diverse clinical situations and improve patient care. Efforts should also be directed toward improving the model's sensitivity to very small infarcts. Techniques such as enhancing the model's spatial resolution capabilities or implementing specialized loss functions that emphasize small infarct detection could be explored. Addressing this limitation will be crucial for ensuring that the model can perform effectively in all clinical scenarios, including those where the identification of small infarcts is vital.

## 6. Conclusions

This study presents a novel deep learning framework for ischemic lesion segmentation in MRI images, specifically tailored to the ISLES 2022 dataset. The proposed a novel model architecture, which integrates DenseNet121 as the encoder, SelfONN in the decoder, and advanced CSCA and DSE blocks, achieved superior segmentation accuracy. A critical contribution of this research is the development of a composite loss function that equalizes the weights of Dice and Jaccard losses, significantly enhancing the model's performance by ensuring a balanced optimization of segmentation metrics. Additionally, our approach leveraged multimodal input images, combining DWI, ADC, and enhanced DWI modalities. This multimodal strategy allowed the model to capture complementary information from different imaging techniques, further improving segmentation accuracy and robustness. The use of these combined modalities proved particularly effective in challenging cases, demonstrating the model's ability to generalize across varying lesion presentations. Our model's ability to beat traditional techniques highlights the possibility of incorporating it into clinical operations, providing a reliable and efficient tool for stroke diagnosis and management. The combination of a novel loss function, advanced architectural components, and multimodal input strategies positions our model as a promising candidate for real-world clinical applications. Future work will focus on refining the model further, expanding its applicability across diverse medical imaging challenges, and promoting the adoption of AI-driven solutions in healthcare, ultimately contributing to improved patient care and advancements in medical imaging.

**Data Availability Statement**
The dataset utilized in this study can be obtained upon request from the corresponding author.

**Conflict of Interest**
The authors have no conflicts of interest to disclose for this study.